\documentclass[preprint,aps,amsmath,showpacs,pre]{revtex4}

\usepackage{graphicx}
\usepackage{amssymb}
\begin{document}
\title{Bethe lattice solution of a model of SAW's with up to 3
  monomers per site and no restriction}  
\author{Tiago J. Oliveira}
\email{tiago@ufv.br}
\affiliation{Departamento de F\'isica, Universidade Federal de Vi\c
  cosa, 36570-000, Vi\c cosa, MG, Brazil} 
\author{J\"urgen F. Stilck}
\email{jstilck@if.uff.br}
\affiliation{Instituto de F\'{\i}sica and National Institute of Science and 
Technology for Complex Systems, Universidade Federal
  Fluminense, Av. Litor\^anea s/n, 24210-346, Niter\'oi, RJ, Brazil} 
\date{\today}

\begin{abstract}
In the multiple monomers per site (MMS) model, polymeric chains are
represented by walks on a lattice which may visit each site up to $K$
times. We have solved the unrestricted version of this model, where
immediate reversals of the walks are allowed (RA) for $K=3$ on a Bethe
lattice with arbitrary coordination number in the grand-canonical
formalism. We found transitions between a non-polymerized and two
polymerized phases, which may be continuous or discontinuous. In the
canonical situation, the transitions between the extended and the
collapsed polymeric phases are always continuous. The transition line
is partly composed by tricritical points and partially by critical
endpoints, both lines meeting at a multicritical point. In the
subspace of the parameter space where the model is related to
SASAW's (self-attracting self-avoiding walks), the collapse transition
is tricritical. We discuss the relation of our results with
simulations and previous Bethe and Husimi lattice calculations for the
MMS model found in the literature.
\end{abstract}


\maketitle

\section{Introduction}
\label{intro}

The thermodynamic behavior of polymers, both in solution or in a melt,
may be studied using continuum or lattice models \cite{f66}. In particular,
linear polymers in lattice models are usually described by self- and
mutually avoiding walks on the lattice (SAW's). The excluded volume
interactions, are essential to reproduce the correct scaling behavior of
the system \cite{dg79}. This constraint also adds considerable
difficulties to the study of the models, when compared to the case of
random walks, where much statistical results are known analytically
\cite{mw79}. As an example of the effect of the self-avoidance
constraint on the asymptotic properties of a single walk on a lattice,
we may recall that, while the size of the region occupied by a random
walk with $\ell$ steps on a lattice, measured by the end-to-end
distance or the radius of gyration, grows as $\ell^{1/2}$ in the limit
$\ell \to \infty$, for lattices of dimension below 4, the asymptotic
behavior for SAW's is described by an exponent which is larger than
$1/2$, and 
thus the size of the region occupied by the walk on these lattices
grows faster with the number of steps of the walk when excluded volume
interactions are present. For two-dimensional lattices this exponent is
known to be equal to $3/4$ \cite{n82}. We notice that since this
exponent is larger than $1/d=1/2$ in this case, the density of
monomers vanishes in the region occupied by the polymer. 

The basic property which describes the behavior of a single self-avoiding
walks on a lattice is the number of walks with $\ell$ steps, starting
from the origin of the lattice. We may consider the walks to be
chains, so that the steps are bonds which link successive monomers of
the polymeric chain. The number of monomers of a chain, which may be
called its molecular weight $M$, is the number
of lattice sites visited by the SAW, so that
$M=\ell+1$. If we wish to study a single chain in the grand-canonical
ensemble, where the number of monomers fluctuates, we
associate a fugacity $z=\exp(\beta \mu)$ to each monomer in the chain,
where $\beta=(k_BT)^{-1}$ and $\mu$ is the chemical potential of a
monomer. The grand-canonical partition function will then be given by:
\begin{equation}
Y(z)=\sum_{M} C_M z^M,
\label{gcpf}
\end{equation}
where $C_M$ is the number of configurations of a chain with $M$
monomers ($M-1$ steps). Alternatively, this partition function may be
viewed as the generating function for the numbers of chain
configurations 
$C_M$. So far, all allowed configurations are associated to the same
energy, and thus the model is athermal. The model defined in this way
displays a phase transition, a non-polymerized phase is
stable at low values of the fugacity $z$, and for fugacities above a
critical value $z_c$ a polymerized phase is stable, with a positive
density of monomers placed on the lattice. At the critical fugacity,
the density of the polymerized phase vanishes, so that the
polymerization transition is continuous. The critical value of the
fugacity is related to the asymptotic behavior of the numbers of SAW's
$C_M$ in the large $M$ limit. There is good numerical evidence that
$C_M \sim M^{\gamma-1}q_e^M$, where the effective coordination number
$q_e$ is smaller than the coordination number of the lattice, and the
critical exponent $\gamma$ is equal to $4/3$ in two dimensions, $7/6$
in three dimensions and 1 in four dimensions or above \cite{d70}. The
effective coordination number is the inverse of the critical fugacity
$q_e=1/z_c$, and it is easy to show that the grand-canonical partition
function Eq. (\ref{gcpf}) is singular at this value of the fugacity, its
asymptotic behavior being given by $Y(z) \sim A(1-q_e z)^\gamma$. In
the canonical ensemble, the system is critical in the thermodynamic
limit $M \to \infty$ \cite{tiago09}. The
recognition that the contributions to the high-temperature series
expansion of the $n$-vector model of magnetism in the limit $n \to 0$
reduce to SAW's on the lattice \cite{dg72} has allowed the application
of renormalization group methods to the polymer transition, linking
this problem to the much studied ferromagnetic transition in the
$n$-vector model.

\begin{figure}[!h]
\centering
\includegraphics[width=9cm]{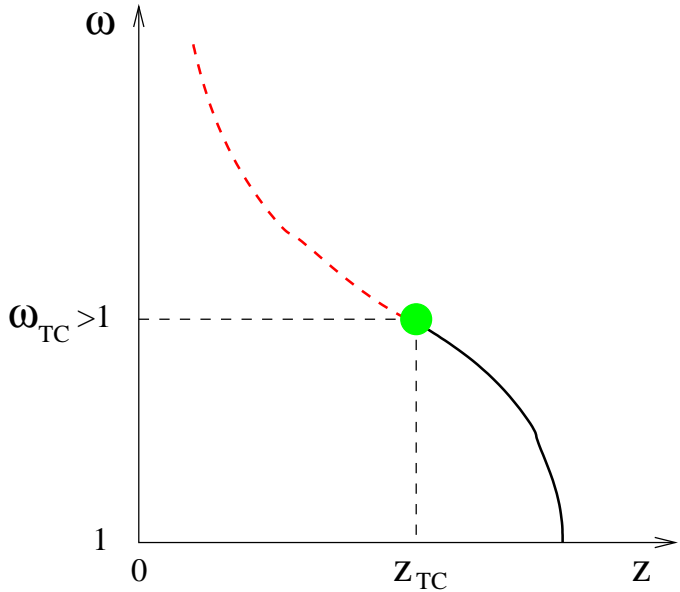}
\caption{(Color online) Schematic phase diagram of the model of
  self-attracting self-avoiding walks (SASAW's) on a lattice. The
  continuous polymerization transition is represented by a full line
  (black on line) and the discontinuous transition is located at the
  dashed line (red on line). Both transition lines are separated by a
  tricritical point, represented by the circle (green on line).}
\label{pdsasaws}
\end{figure}

The athermal polymerization model may be generalized by including
attractive interactions between monomers located on first neighbor
sites and which are not connected by a polymer bond. This model of
self-avoiding self-attracting walks (SASAW's) is usually used as an
effective model to study the behavior of a polymer in a poor solution,
the attractive interaction mimics the energetically unfavorable
contact between solvent molecules and polymeric monomers
\cite{dg79}. Now, besides the monomer fugacity, an additional
parameter is present in the model, the Boltzmann weight
$\omega=\exp(\beta\epsilon)$, where $-\epsilon$ is the energy
associated to each monomer-monomer interaction, and there
will be a competition between the repulsive excluded volume
interactions and the attractive interactions. The model reduces to
the previous one for $\omega=1$, and as $\omega$ is increased the
continuous polymerization transition happens at lower values of the
fugacity $z$, becoming discontinuous if $\omega$ exceeds a value
$\omega_{TC}$. Thus, a tricritical point is found in the phase diagram
of the model at $(z_{TC},\omega_{TC})$, as may be seen in the
schematic diagram shown in Fig. \ref{pdsasaws}. In the canonical
situation, the system is always on the border between the
non-polymerized and the polymerized phase \cite{tiago09}. At high
temperatures (low values of $\omega$), the
polymerized phase is indistinguishable from the non-polymerized phase,
and thus has vanishing monomer density. This phase is sometimes called
coil phase in the polymer physics literature. Below the tricritical
temperature (called $\theta$ point \cite{f66}), the chain
is collapsed and the 
polymerized phase has nonzero density (globule phase). Again it is
possible to map the SASAW's model on a generalized ferromagnetic
$n$-vector model \cite{dg75}. In two dimensions, the tricritical point
of the model has been studied in detail using transfer matrix
techniques \cite{ds85}, and the tricritical value of the exponent
which characterizes the scaling behavior of the radius of
gyration is $\nu_t=4/7$ \cite{ds87}.

More recently, an alternative model has been proposed by Krawczyk et
al \cite{kpor06} for the collapse transition of polymers. In this
model, which we may call MMS (multiple monomers per site) model and
is a generalization of the Domb-Joyce model 
\cite{dj72}, up to $K$ monomers may occupy the same site of the
lattice.  The canonical version of the model, with chains of fixed
(large) number of monomers $M$, was
studied for $K=3$ using computer simulations on the square and cubic
lattices. Besides the case with no additional restrictions, which was
named RA (immediate reversals allowed) model by the authors, a more
restrictive model, where the chain is not allowed to return to the
lattice site it occupied two steps ago, (RF model), was also
studied. Collapse transitions were found only for the RF model on the
cubic lattice, indicating that, at least for this lattice, the
restrictions seem to be essential for the existence of these
transition. The weight of a site with two and three monomers in the
model is $\omega_1$ and $\omega_2$, respectively, and in the two-dimensional
parameter space of the model defined by the variables $\beta_1=\ln
\omega_1$ and $\beta_2=\ln \omega_2$, it seems that the
collapsed polymerized phase (globule)  is separated from the regular
polymerized extended phase (coil) by lines of continuous and
discontinuous transitions, both transition lines meeting at a
tricritical point. We note that, in the SASAW's model discussed above,
the extended-collapsed transition in the canonical situation is
continuous and of tricritical nature. One point which needs to be
understood is the apparent absence of transitions in both models on
the square lattice and in the RA model on the cubic lattice.

Motivated by the questions above, the grand-canonical version of the
MMS model was studied on Bethe and Husimi
lattices. Initially, both versions of the model (RA and RF) with $K=2$
were solved on a Bethe lattice with general coordination number $q$
\cite{pablo07}. Although these initial calculation for the models
resulted in phase diagrams with  some qualitative differences when
compared to the usual behavior of SASAW's, a revision using a
different and better fundamented procedure to find the coexistence
loci resulted in 
diagrams which are similar to the ones for SASAW's, in both cases (RA
and RF), with continuous transitions between the
polymerized phases in the canonical formalism \cite{ss10}. The
solution of the $K=2$ RF model on the Husimi lattice \cite{oss08}
lead to a phase diagram similar to the one found for the same model
on the Bethe lattice. A natural interpretation of this model is to
consider that monomers on the same lattice attract each other, so that
the statistical weights of sites with one and two monomers will be
$\omega_1=z$ and $\omega_2=\omega z^2$, where $z$ is the fugacity of a
monomer and $\omega=\exp(-\beta \epsilon)$ is the Boltzmann factor
associated to the attractive interaction energy $\epsilon$ between
monomers on the same site. While the tricritical point in the Bethe
lattice solution of the model corresponds to $\omega_{TC}=1$, the solution
on the Husimi lattice shows the tricritical point located at $\omega
\approx 1.09$, in the 
region of attractive monomer-monomer interactions, as expected. More
recently, the RF model for $K=3$ was solved on the Bethe lattice
in the grand-canonical ensemble \cite{tiago09}. In the
two-dimensional 
subspace of the three-dimensional parameter space used in the
grand-canonical calculations, which corresponds to the canonical phase
diagram, again the transitions between the polymerized phases are
always continuous. Two transition lines, one composed by tricritical
points and the other by critical endpoints, meet at a multicritical
point, not far from the region in the parameter space where the
tricritical point was found in the original simulations of the $K=3$
RF model on the cubic lattice.

Here we present calculations for the $K=3$ RA model on the Bethe
lattice, partially motivated by the surprising result in the
original simulations that no transition was found for this
unrestricted model between the polymerized phases, while at least in
the $K=2$ case the Bethe lattice calculations revealed no qualitative
differences in the phase diagrams of the RA and RF models, both
similar to the one found for the SASAW's model. In section
\ref{defmod} we define the model in more detail and present its
solution on the Bethe lattice in terms of recursion relations. The
thermodynamic behavior of the model is determined by the fixed points
of the recursion relations, together with a bulk free energy which is
useful to locate the coexistence loci, and these results may be found
in section \ref{tpm}. Final discussions and the conclusion are
presented in section \ref{conclusao}.

\section{Definition of the model and solution in terms of recursion
  relations} 
\label{defmod}

We study the MMS-RA model proposed by Krawczyk et al in \cite{kpor06} in
the core of a
Cayley tree with arbitrary coordination number $q$. In this model,
self- and mutually avoiding walks are considered but the excluded
volume 
condition is relaxed, so that each site of the tree may be occupied by
up to $K=3$ monomers, or, equivalently, each lattice site way be
visited up to three times by the walks. No other restriction is imposed
in the model, so that immediate reversals of the walk are allowed (RA model),
differently from the more restrictive model studied in 
\cite{tiago09} where immediate reversals are forbidden (RF model), and
thus a subset of the configurations of the walks considered here was
included.  

\begin{figure}[!h]
\centering
\includegraphics[width=9cm]{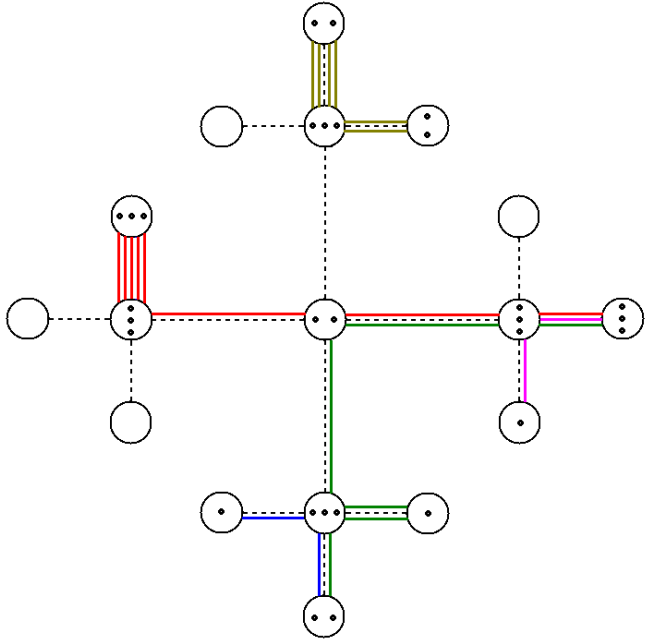}
\caption{(Color online) A contribution to the partition function of
  the model on a Cayley tree with $q = 4$ and 3 generations. The
  weight of this contribution will be $\omega_{1}^{3} \omega_{2}^{4}
  \omega_{3}^{6}$.} 
\label{rede}
\end{figure}

As usual, the endpoints of the walks are placed on the surface of the
tree. Like in the original model \cite{kpor06}, a walk is described by
the sequence of sites that it visits, so that the monomers placed on
the same site are considered to be
indistinguishable. A statistical weight $\omega_{i}$ is associated to
a site occupied by $i$ monomers, with $i=1,2,3$. So, the
grand-canonical partition function of the model will be given by: 
\begin{equation}
 Y = \sum_{N_{1},N_{2},N_{3}} \omega_{1}^{N_{1}} \omega_{2}^{N_{2}}
 \omega_{3}^{N_{3}} 
\end{equation}
where the sum is over the configurations of the walks on the tree,
while $N_{i}$ , $i = 1, 2, 3$, is the number of sites visited $i$
times by the walks. In Fig. \ref{rede}, an example of a
Cayley tree with three generations of sites is shown, as well as the
contribution to the partition 
function which corresponds to the configuration of the walks in this
case.  

\begin{figure}[!h]
\centering
\includegraphics[width=8cm]{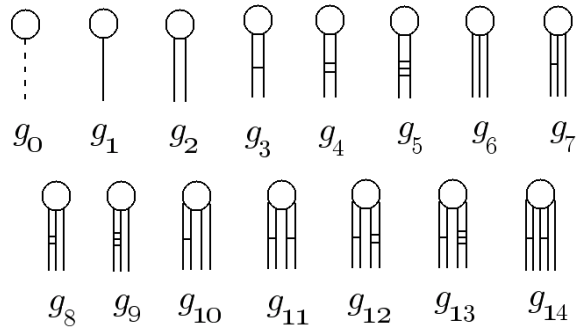}
\caption{Illustration of the rooted subtrees which correspond to the
  partial partition functions. The meaning of the horizontal lines
  between bonds on the root edge is discussed in the text.} 
\label{roots}
\end{figure}

To solve the model on the Bethe lattice (the core of the Cayley tree)
we start considering rooted subtrees, defining partial partition
functions (ppf's) for them, where we sum over all possible
configurations of the chains for
a fixed configuration of the root of the subtree (this is the reason
for calling the partition functions partial). We thus define
fourteen partial partition functions $g_{i}$, $i = 0, 1, \ldots, 14$,
shown in Fig. \ref{roots}. The number of partial partition functions
we need to define for the RA model is larger than the one we used for
the RF  model, where four root configurations were sufficient
\cite{tiago09}, since more configurations are allowed in the present case. 

When we define the possible root configurations, with up to four
polymer bonds on the root edge. It is important to
notice that, since immediate reversals of the walk are now
allowed, it is possible to have closed loops on the tree and the
ppf's have to be carefully defined in order to avoid rings in
the walks. This possibility does not exist in the RF model on the Bethe
lattice. When rings are allowed, even the universality class of a
polymer model changes to a model with $n=1$ components in the order
parameter, that is, to the universality class of the Ising
model. Therefore, in ppf's with two or more bonds on the root edge, we
need to  
distinguish between bond pairs that are connected (in earlier
generations of the tree), or not. In Fig.
\ref{roots}, we have four ppf's with two bonds in the root site
($g_{2}$ to $g_{5}$), for example. In bond pairs not connected by
horizontal lines in our notation,
such as in the root configurations for $g_{2}$ and $g_{6}$ (taken two by
two) and one pair of $g_{10}$, 
the two walks will never meet on the tree at a site in earlier
generations, or, in other words, if the walks are followed all the way
to the surface of the tree, they end at different surface sites. When
we draw horizontal lines connecting two
bonds at the root edge, it means that they are connected to the same
site in earlier generations and there are three
possibilities for this: 1) The two bonds are connected to the same
monomer of the root site (one line); 2) Both walks are connected to
the same site in some
earlier generation and visit the same sites of the tree (two lines);
and 3) Same as case 2, but the walks visit different sites of the tree
(three 
lines). In the last case, the two \textit{bonds} are distinguishable,
because the sequence of visited sites is different if we begin in one
or other bond, although the visited sites are the same. These
definitions are applied for every ppf with at least two bonds in the
root site, and leads to the rather large number of ppf's we need to
define.

We then proceed obtaining recursion relations for the ppf's, by
considering the operation of attaching $q-1$ 
subtrees with a certain number of generations to a new root site and
edge, thus building a subtree with an additional generation. Below the
recursion relations are presented. In general, the partial partition
function  $g'_{i}$ with an additional generation is the sum of
contributions involving the parameters of the model and partial
partition functions $g_i$. The primes denote the partial partition
function of the subtree with one more generation. Whenever
appropriate, the contributions to the sums begin with a product of two
numerical factors, the first of which is the multiplicity of the
configuration of the incoming bonds and the second is the multiplicity
of the connections with the monomers located at the new root site. In the
expressions below, $f_{i} \equiv \binom{\sigma}{i}
g_{0}^{i}$, where $\sigma=q-1$ is the ramification of the tree. The
recursion relations for the 14 ppf's are:  

\begin{eqnarray}
g'_{0}&=&f_{0} + \omega_{1} [ f_{2} g_{1}^2 + f_{1} g_{2} ]  +
\omega_{2} [f_{4} \times 3 g_{1}^4 + 3 f_{3} \times 3 g_{1}^2 g_{2} +
3 f_{3} g_{1}^2 (g_{3} + g_{4} + 2 g_{5}) + 2 f_{2} \times 3 g_{1}
g_{6} + \nonumber \\ 
&& 2 f_{2} g_{1} (g_{7} + g_{8} + 2 g_{9}) + f_{2} \times 3 g_{2}^2 +
2 f_{2} g_{2} (g_{3} + g_{4} + 2 g_{5}) + f_{1} g_{10} ]  +
\omega_{3} [f_{6} \times 15 g_{1}^6 + \nonumber \\ 
&& 5 f_{5} \times 15 g_{1}^4 g_{2} + 6 f_{4} \times 15 g_{1}^2 g_{2}^2
+ 12 f_{4} \times 6 g_{1}^2 g_{2} (g_{3} + g_{4} + 2 g_{5}) + 6 f_{3}
\times 15 g_{1} g_{2} g_{6} + \nonumber \\ 
&& 6 f_{3} \times 6 g_{1} g_{2} (g_{7} + g_{8} + 2 g_{9}) + 5 f_{5}
\times 6 g_{1}^4 (g_{3} + g_{4} + 2 g_{5}) + 6 f_{4} \times 2 g_{1}^2
(g_{3} + g_{4} + 2 g_{5})^2 + \nonumber \\ 
&& 6 f_{3} \times 6 g_{1} (g_{3} + g_{4} + 2 g_{5}) g_{6} + 6 f_{3}
\times 2 g_{1} (g_{3} + g_{4} + 2 g_{5}) (g_{7} + g_{8} + 2 g_{9}) + 4
f_{4} \times 15 g_{1}^3 g_{6} + \nonumber \\ 
&& 4 f_{4} \times 6 g_{1}^3 (g_{7} + g_{8} + 2 g_{9}) + 3 f_{3} \times
6 g_{1}^2 g_{10} + 3 f_{3} g_{1}^2 (g_{11} + 2 g_{12} + 4 g_{13}) + 2
f_{2} g_{1} g_{14} + f_{3} \times 15 g_{2}^3 + \nonumber \\ 
&& 3 f_{3} \times 6 g_{2}^2 (g_{3} + g_{4} + 2 g_{5}) + 3 f_{3} \times
2 g_{2} (g_{3} + g_{4} + 2 g_{5})^2 + 2 f_{2} \times 6 g_{2} g_{10} +
2 f_{2} g_{2} (g_{11} + 2 g_{12} + 4 g_{13}) + \nonumber \\ 
&& 2 f_{2} \times 2 (g_{3} + g_{4} + 2 g_{5}) g_{10} + f_{2} \times 15
g_{6}^2 + 2 f_{2} \times 6 g_{6} (g_{7} + g_{8} + 2 g_{9}) + f_{2}
\times 2 (g_{7} + g_{8} + 2 g_{9})^2 ];
\end{eqnarray}

\begin{eqnarray}
g'_{1}&=&\omega_{1} f_{1} g_{1}  +  \omega_{2} [f_{3} \times 3 g_{1}^3
+ 2 f_{2} \times 3 g_{1} g_{2} + 2 f_{2} g_{1} (g_{3} + g_{4} + 2
g_{5}) + f_{1} \times 3 g_{6} + f_{1} (g_{7} + g_{8} + 2 g_{9})]  +
\nonumber \\ 
&& \omega_{3} [f_{5} \times 15 g_{1}^5 + 4 f_{4} \times 15 g_{1}^3
g_{2} + 3 f_{3} \times 15 g_{1} g_{2}^2 + 6 f_{3} \times 6 g_{1} g_{2}
(g_{3} + g_{4} + 2 g_{5}) +  \nonumber \\ 
&& 4 f_{4} \times 6 g_{1}^3 (g_{3} + g_{4} + 2 g_{5})+3 f_{3} \times 2
g_{1} (g_{3} + g_{4} + 2 g_{5})^2 + 3 f_{3} \times 15 g_{1}^2 g_{6} +
\nonumber \\ 
&& 3 f_{3} \times 6 g_{1}^2 (g_{7} + g_{8} + 2 g_{9}) + 2 f_{2} \times
6 g_{1} g_{10}+2 f_{2} g_{1} (g_{11} + 2 g_{12} + 4 g_{13}) + 2 f_{2}
\times 15 g_{2} g_{6} +  \nonumber \\ 
&& 2 f_{2} \times 6 g_{2} (g_{7} + g_{8} + 2 g_{9}) + 2 f_{2} \times 6
(g_{3} + g_{4} + 2 g_{5}) g_{6} + 2 f_{2} \times 2 (g_{3} + g_{4} + 2
g_{5}) (g_{7} + g_{8} + 2 g_{9}) +  \nonumber \\ 
&& f_{1} g_{14}];
\end{eqnarray}

\begin{eqnarray}
g'_{2}&=&\omega_{2} [f_{2} g_{1}^2 + f_{1} g_{2}]  +  \omega_{3}
[f_{4} \times 6 g_{1}^4 + 3 f_{3} \times 6 g_{1}^2 g_{2} + 3 f_{3}
\times 2 g_{1}^2 (g_{3} + g_{4} + 2 g_{5}) + 2 f_{2} \times 6 g_{1}
g_{6} + \nonumber \\ 
&& 2 f_{2} \times 2 g_{1} (g_{7} + g_{8} + 2 g_{9}) + f_{2} \times 6
g_{2}^2 + 2 f_{2} \times 2 g_{2} (g_{3} + g_{4} + 2 g_{5}) + f_{1}
\times 2 g_{10}];
\end{eqnarray}

\begin{eqnarray}
g'_{3}&=&\omega_{1}  +  \omega_{2} [f_{2} g_{1}^2 + f_{1} g_{2}]  +
\omega_{3} [f_{4} \times 3 g_{1}^4 + 3 f_{3} \times 3 g_{1}^2 g_{2} +
3 f_{3} g_{1}^2 (g_{3} + g_{4} + 2 g_{5}) + 2 f_{2} \times 3 g_{1}
g_{6} + \nonumber \\ 
&& 2 f_{2} g_{1} (g_{7} + g_{8} + 2 g_{9}) + f_{2} \times 3 g_{2}^2 +
2 f_{2} g_{2} (g_{3} + g_{4} + 2 g_{5}) + f_{1} g_{10}];
\end{eqnarray}

\begin{eqnarray}
g'_{4}&=&\omega_{2} f_{1} (g_{3} + g_{4})  +  \omega_{3} [3 f_{3}
g_{1}^2 (g_{3} + g_{4}) + 2 f_{2} g_{1} (g_{7} + g_{8}) + 2 f_{2}
g_{2} (g_{3} + g_{4}) + f_{1} g_{10} + f_{1} g_{11}];
\end{eqnarray}

\begin{eqnarray}
g'_{5}&=&\omega_{2} f_{1} g_{5} + \omega_{3} [3 f_{3} g_{1}^2 g_{5} + 2
f_{2} g_{1} g_{9} + 2 f_{2} g_{2} g_{5} + f_{2} (g_{3} + g_{4} + 2
g_{5})^2 + f_{1} (g_{12} + 2 g_{13})];
\end{eqnarray}

\begin{eqnarray}
g'_{6}&=&\omega_{3} [f_{3} g_{1}^3 + 2 f_{2} g_{1} g_{2} + f_{1}
  g_{6}];
\end{eqnarray}

\begin{eqnarray}
g'_{7}&=&\omega_{2} f_{1} g_{1}  +  \omega_{3} [f_{3} \times 3 g_{1}^3 +
2 f_{2} \times 3 g_{1} g_{2} + 2 f_{2} g_{1} (g_{3} + g_{4} + 2 g_{5})
+ f_{1} \times 3 g_{6} + f_{1} (g_{7} + g_{8} + 2 g_{9})];
\end{eqnarray}

\begin{eqnarray}
g'_{8}&=&\omega_{3} [2 f_{2} g_{1} (g_{3} + g_{4}) + f_{1} (g_{7} +
g_{8})];
\end{eqnarray}

\begin{eqnarray}
g'_{9}&=&\omega_{3} [2 f_{2} g_{1} g_{5} + f_{1} g_{9}];
\end{eqnarray}

\begin{eqnarray}
g'_{10}&=&\omega_{3} [f_{2} g_{1}^2 + f_{1} g_{2}];
\end{eqnarray}

\begin{eqnarray}
g'_{11}&=&\omega_{2} + \omega_{3} [f_{2} g_{1}^2 + f_{1} g_{2}];
\end{eqnarray}

\begin{eqnarray}
g'_{12}&=&\omega_{3} f_{1} (g_{3} + g_{4});
\end{eqnarray}

\begin{eqnarray}
g'_{13}&=&\omega_{3} f_{1} g_{5};
\end{eqnarray}

\begin{eqnarray}
g'_{14}&=&\omega_{3} f_{1} g_{1}.
\end{eqnarray}

The partial partition functions are expected to grow exponentially with the
number of iterations, so we define ratios of them, which usually
remain finite in the thermodynamic 
limit. Furthermore, we notice in the above equations that some
ppf's only appear in sums, they are $(g_{3}+g_{4}+2 g_{5})$,
$(g_{7}+g_{8}+2 g_{9})$ and $(g_{11}+2 g_{12}+4 g_{13})$. Thus, it is
convenient to define the following ratios: 
\begin{eqnarray}
R_{1}=\dfrac{g_{1}}{g_{0}} , \quad R_{2}=\dfrac{g_{2}}{g_{0}} , \quad
R_{3}=\dfrac{(g_{3}+g_{4}+2 g_{5})}{g_{0}} , \quad
R_{4}=\dfrac{g_{6}}{g_{0}} \nonumber \\ 
R_{5}=\dfrac{(g_{7}+g_{8}+2 g_{9})}{g_{0}} , \quad
R_{6}=\dfrac{g_{10}}{g_{0}} , \quad R_{7}=\dfrac{(g_{11}+2 g_{12}+4
  g_{13})}{g_{0}} \quad \text{and} \quad R_{8}=\dfrac{g_{14}}{g_{0}}.  
\end{eqnarray}

From the recursion relations for the ppf's, similar expressions may be
obtained for the ratios. Denoting the binomial coefficients as $b_{i}
\equiv \binom{\sigma}{i}$, the recursion relations for the ratios are:

\begin{eqnarray}
R'_{1}&=& \frac{1}{D}[\omega_{1} b_{1} R_{1}  +  \omega_{2} (3 b_{3}
R_{1}^3 + 6 b_{2} R_{1} R_{2} + 2 b_{2} R_{1} R_{3} + 3 b_{1} R_{4}+
b_{1} R_{5})  + \nonumber \\ 
&& \omega_{3} (15 b_{5} R_{1}^5 + 60 b_{4} R_{1}^3 R_{2} + 45 b_{3}
R_{1} R_{2}^2 + 36 b_{3} R_{1} R_{2} R_{3} + 24 b_{4} R_{1}^3 R_{3}+
\nonumber \\ 
&& 6 b_{3} R_{1} R_{3}^2 + 45 b_{3} R_{1}^2 R_{4}+ 18 b_{3} R_{1}^2
R_{5} + 12 b_{2} R_{1} R_{6}+2 b_{2} R_{1} R_{7} + 30 b_{2} R_{2}
R_{4}+  \nonumber \\ 
&& 12 b_{2} R_{2} R_{5} + 12 b_{2} R_{3} R_{4}+ 4 b_{2} R_{3} R_{5} +
b_{1} R_{8})];
\label{rr1}
\end{eqnarray}

\begin{eqnarray}
R'_{2}&=&\frac{1}{D} [\omega_{2} (b_{2} R_{1}^2 + b_{1} R_{2})  +
\omega_{3} (6 b_{4} R_{1}^4 + 18 b_{3} R_{1}^2 R_{2} + 6 b_{3} R_{1}^2
R_{3} + 12 b_{2} R_{1} R_{4}+ \nonumber \\ 
&& 4 b_{2} R_{1} R_{5} + 6 b_{2} R_{2}^2 + 4 b_{2} R_{2} R_{3} + 2
b_{1} R_{6})];
\label{rr2}
\end{eqnarray}

\begin{eqnarray}
R'_{3}&=& \frac{1}{D}[\omega_{1}  +  \omega_{2} (b_{2} R_{1}^2 + b_{1}
R_{2} + b_{1} R_{3})  +  \omega_{3} (3 b_{4} R_{1}^4 + 9 b_{3} R_{1}^2
R_{2} + 6 b_{3} R_{1}^2 R_{3} + \nonumber \\ 
&& 6 b_{2} R_{1} R_{4}+ 4 b_{2} R_{1} R_{5} + 3 b_{2} R_{2}^2 + 4
b_{2} R_{2} R_{3} + 2 b_{2} R_{3}^2 + 2 b_{1} R_{6} + b_{1} R_{7})];
\label{rr3}
\end{eqnarray}

\begin{eqnarray}
R'_{4}&=&\frac{\omega_{3}}{D} [b_{3} R_{1}^3 + 2 b_{2} R_{1} R_{2} +
b_{1} R_{4}];
\label{rr4}
\end{eqnarray}

\begin{eqnarray}
R'_{5}&=&\frac{1}{D}[\omega_{2} b_{1} R_{1}  +  \omega_{3} (3 b_{3}
R_{1}^3 + 6 b_{2} R_{1} R_{2} + 4 b_{2} R_{1} R_{3} + 3 b_{1} R_{4}+ 2
b_{1} R_{5})];
\label{rr5}
\end{eqnarray}

\begin{eqnarray}
R'_{6}&=&\frac{\omega_{3}}{D} [b_{2} R_{1}^2 + b_{1} R_{2}];
\label{rr6}
\end{eqnarray}

\begin{eqnarray}
R'_{7}&=&\frac{1}{D} [\omega_{2} + \omega_{3} (b_{2} R_{1}^2 + b_{1}
R_{2} + 2 b_{1} R_{3})];
\label{rr7}
\end{eqnarray}

\begin{eqnarray}
R'_{8}&=&\frac{\omega_{3}}{D} b_{1} R_{1}.
\label{rr8}
\end{eqnarray}

The denominator $D$ is defined as:
\begin{eqnarray}
D&=&1 + \omega_{1} ( b_{2} R_{1}^2 + b_{1} R_{2} )  +  \omega_{2} (3
b_{4} R_{1}^4 + 9 b_{3} R_{1}^2 R_{2} + 3 b_{3} R_{1}^2 R_{3} + 6
b_{2} R_{1} R_{4}+2 b_{2} R_{1} R_{5} + \nonumber \\ 
&&  3 b_{2} R_{2}^2 + 2 b_{2} R_{2} R_{3} + b_{1} R_{6} )  +
\omega_{3} (15 b_{6} R_{1}^6 + 75 b_{5} R_{1}^4 R_{2} + 90 b_{4}
R_{1}^2 R_{2}^2 + 72 b_{4} R_{1}^2 R_{2} R_{3} +  \nonumber \\ 
&& 36 b_{3} R_{1} R_{3} R_{4}+90 b_{3} R_{1} R_{2} R_{4}+36 b_{3}
R_{1} R_{2} R_{5} + 30 b_{5} R_{1}^4 R_{3} + 12 b_{4} R_{1}^2 R_{3}^2
+ 12 b_{3} R_{1} R_{3} R_{5} + \nonumber \\ 
&& 60 b_{4} R_{1}^3 R_{4}+ 24 b_{4} R_{1}^3 R_{5} + 18 b_{3} R_{1}^2
R_{6} + 3 b_{3} R_{1}^2 R_{7} + 2 b_{2} R_{1} R_{8} + 15 b_{3} R_{2}^3
+ 18 b_{3} R_{2}^2 R_{3} + \nonumber \\ 
&& 6 b_{3} R_{2} R_{3}^2 + 12 b_{2} R_{2} R_{6} + 2 b_{2} R_{2} R_{7}
+ 4 b_{2} R_{3} R_{6} + 15 b_{2} R_{4}^2 + 12 b_{2} R_{4}R_{5} + 2
b_{2} R_{5}^2 ).
\end{eqnarray}

The grand-canonical partition function of the model on the Cayley tree
may be obtained 
if we consider the operation of attaching $q$ subtrees to the central
site of the lattice, similar to the one used for deriving the
recursion relations for the ppf's. The result will be: 
\begin{equation}
 Y = g_{0}^{q} (1+P+Q+S),
\label{gcpf1}
\end{equation}
where:
\begin{subequations}
\begin{eqnarray}
P &=& \omega_{1} ( c_{2} R_{1}^2 + c_{1} R_{2} ); \\
Q &=& \omega_{2} (3 c_{4} R_{1}^4 + 9 c_{3} R_{1}^2 R_{2} + 3 c_{3}
R_{1}^2 R_{3} + 6 c_{2} R_{1} R_{4}+2 c_{2} R_{1} R_{5} + \nonumber \\ 
&&  3 c_{2} R_{2}^2 + 2 c_{2} R_{2} R_{3} + c_{1} R_{6} );\\
S &=& \omega_{3} (15 c_{6} R_{1}^6 + 75 c_{5} R_{1}^4 R_{2} + 90 c_{4}
R_{1}^2 R_{2}^2 + 72 c_{4} R_{1}^2 R_{2} R_{3} + 36 c_{3} R_{1} R_{3}
R_{4} +  \nonumber \\ 
&& 90 c_{3} R_{1} R_{2} R_{4}+36 c_{3} R_{1} R_{2} R_{5} + 30 c_{5}
R_{1}^4 R_{3} + 12 c_{4} R_{1}^2 R_{3}^2 + 12 c_{3} R_{1} R_{3} R_{5}
+ \nonumber \\ 
&& 60 c_{4} R_{1}^3 R_{4}+ 24 c_{4} R_{1}^3 R_{5} + 18 c_{3} R_{1}^2
R_{6} + 3 c_{3} R_{1}^2 R_{7} + 2 c_{2} R_{1} R_{8} + 15 c_{3} R_{2}^3
+  \nonumber \\ 
&& 18 c_{3} R_{2}^2 R_{3} + 6 c_{3} R_{2} R_{3}^2 + 12 c_{2} R_{2}
R_{6} + 2 c_{2} R_{2} R_{7} + 4 c_{2} R_{3} R_{6} + 15 c_{2} R_{4}^2 +
\nonumber \\ 
&& 12 c_{2} R_{4}R_{5} + 2 c_{2} R_{5}^2 ).
\end{eqnarray}
\label{pqrf}
\end{subequations}
where $c_i \equiv \binom{q}{i}$. We notice that the contributions to
$P$, $Q$, $S$ correspond to placing one, two and three monomers on the
central site, respectively.
Using the expressions above, we may obtain the densities at the
central site of the tree, considering the configuration of this site
for each contribution to the grand-canonical partition function
\ref{gcpf1}.  The density of sites occupied by one
($\rho_{1}$), two ($\rho_{2}$) and three ($\rho_{3}$) monomers are
given, respectively, by: 
\begin{subequations}
\begin{eqnarray}
\rho_{1}&=&\frac{P}{1\,+\,P\,+\,Q\,+\,S}, \\
\rho_{2}&=&\frac{Q}{1\,+\,P\,+\,Q\,+\,S}, \mbox{and} \\
\rho_{3}&=&\frac{S}{1\,+\,P\,+\,Q\,+\,S} .
\end{eqnarray} 
\label{den}
\end{subequations}
The Bethe lattice solution of the model is defined by its
thermodynamic behavior in the core of the tree, represented by the
densities just defined.
The total density of monomers on the Bethe lattice, that is, the total
number of monomers divided by the number of sites, is $\rho= \rho_{1}+2
\rho_{2}+3\rho_{3}$, and will be in the range $0 \le \rho \le 3$.

\section{Thermodynamic properties of the model}
\label{tpm}

The thermodynamic phases of the system on the Bethe lattice will be
given by the stable fixed points of the recursion relations, which are
reached after infinite iterations of the recursion relations and thus
correspond to the thermodynamic limit. We find three different
stable solutions for the fixed
point equations $R_i'=R_i$, associated to one non-polymerized phase
(NP) and two polymerized ones (P1 and P2).

The NP phase is characterized by the fixed point $R_{i} = 0$ for all
$i$, excluding $i = 3$ and $i = 7$. These last two may be obtained
solving the equations: 

\begin{equation}
 2 b_{2} \omega_{3} R_{3}^{2} + (b_{1} \omega_{2} + 2 b_{1}^{2}
 \omega_{3}^{2} -1) R_{3} + (\omega_{1} + b_{1} \omega_{2} \omega_{3})
 = 0,
\label{npeq1}
\end{equation}
 and
\begin{equation}
 R_{7} = \omega_{2} + 2 b_{1} \omega_{3} R_{3} .
\label{npeq2}
\end{equation}

The quadratic equation for $R_{3}$ can be easily solved, but the
explicit expression for $R^{NP}_{3}$ is too large to be shown
here. Looking the equations \ref{pqrf} and \ref{den} we see that
$\rho=0$ in the NP phase, as expected. 

In the two polymerized phases all ratios are non-vanishing and, in
order to obtain the fixed point values, we have to iterate the
recursion relations or solve the fixed point equations numerically.
It is important to remark that the metastable phases with double e
triple occupation of sites, that appear in the RF model
\cite{tiago09}, are absent here. As is  discussed in \cite{pablo07}, the
immediate reversal of the walk makes the probabilities of find these
configurations in the central site vanish. 

The stability limits of all phases are obtained calculating the
jacobian of the recursion relations: 
\begin{equation}
J_{i,j}=\left( \frac{\partial R'_i}{\partial R_j}\right).
\end{equation}
A fixed point is stable if the dominant eigenvalue of the jacobian has
a modulus smaller than unity,
and the stability limit of the corresponding phase (spinodal) is
located in the 
loci where this modulus becomes equal to unity.  

In order to find the coexistence surfaces in the phase diagrams, we
obtain the free energies of the thermodynamic phases of the
model. This free energy may not be calculated directly from the
partition function Eq. (\ref{gcpf1}), since it refers to the whole
Cayley tree and, remembering that in the thermodynamic limit the
number of surface sites correspond to a finite fraction of the total
number of sites, reflects the influence of the surface. The free
energy per site in the core of the tree, which corresponds to the
Bethe lattice solution, may be calculated following the Gujrati's
argument \cite{g95}, which was also derived in another way in
\cite{tiago09}. The result for the grand-canonical free energy per
site on the Bethe lattice (divided by $k_BT$) is:   
\begin{equation}
 \phi_{b} = - \frac{1}{2} \left[ q \ln \left( D \right) - (q-2) \ln
   \left( 1 + P + Q + S \right) \right] 
\end{equation}
Using the spinodals to find the continuous transitions and the free
energy to determine the coexistence surfaces we built the whole phase
diagram of the system. Before presenting the complete
three-dimensional phase diagram, in the space defined by the
statistical weights $\omega_1$, $\omega_2$, and $\omega_3$, it is
instructive look at the thermodynamic behavior
in some cuts of the parameter space. All results presented below
are for a lattice of coordination $q=4$ and qualitatively identical
diagrams are obtained for other values of $q>2$. 

\begin{figure}[!h]
\centering
\includegraphics[width=8cm]{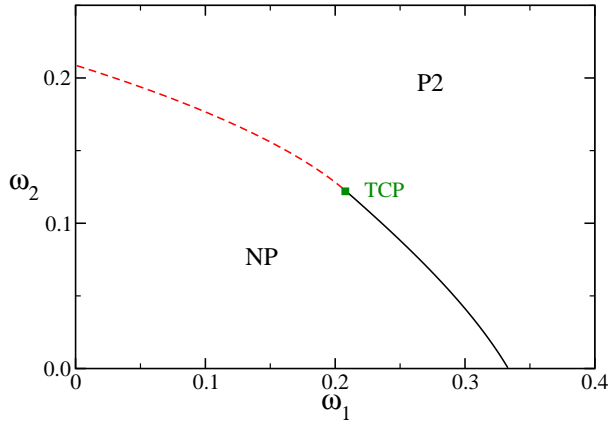}
\caption{(Color online) Phase diagram for $\omega_{3}=0$. The red
  curve (above the tricritical point-TCP) is a first order transition
  and the black line 
  (below the TCP) is a continuous transition between the NP and P2
  phases. The TCP is indicated by the green square.} 
\label{w1w2}
\end{figure}

The diagram for $\omega_{3}=0$ ($K=2$) is shown in Fig. \ref{w1w2}. For 
small values of $\omega_{2}$, we find a continuous transition, between
the NP e P2 phases, which ends up at a tricritical point (TCP) located
at $\omega_{1}^{TCP (K=2)}=0.2177$ and $\omega_{2}^{TCP (K=2)}=0.1146$
(for $q=4$). Above the tricritical point the transition becomes
discontinuous. It is important to stress that this particular case
($\omega_{3}=0$) was studied by one of the authors in \cite{pablo07},
considering distinguishable monomers and using the natural initial
conditions method to find the coexistence lines. There, only a
discontinuous transition was found between the non-polymerized phase
and a polymerized one (called P2 here). We notice that the diagram of
\cite{pablo07} changes if Gujrati's prescription is used to obtain the
bulk free energy and the coexistence lines are determined using this
free energy. Since this latter procedure has a better fundamentation
than the earlier based on {\em natural} initial conditions
\cite{tiago09}, the results provided by it are more reliable.  We
also notice that the phase diagram found here is very similar to one
obtained for the RF model in \cite{tiago09}. However, the RF
tricritical point was located at $\omega^{RF}_{1}=1/3$ and
$\omega^{RF}_{2}=1/9$ \cite{tiago09}, which is far from the location
found here. A generalization of the model for $K=2$, with the RA and
RF models as particular cases, shows a line of tricritical points
joining the two ones for both models \cite{ss10}.

For increasing values of $\omega_{3}$, the qualitative behavior of the
phase diagrams, in the $(\omega_{1},\omega_{2})$ plane, is similar to
the one depicted in
Fig. \ref{w1w2}. Thus, there is a tricritical point (TCP) line in
the three dimensional phase diagram. This line ends up at a
multicritical point (MCP), located at ($\omega_{1}^{MCP}=0.2700819$,
$\omega_{2}^{MCP}=0.0182769$, $\omega_{3}^{MCP}=0.0450771$), so the
TCP line lies in the region $0 \leq \omega_{3} <
\omega_{3}^{MCP}$. For $\omega_{3}>\omega_{3}^{MCP}$, more complex
diagrams are found, as will be discussed below. The location of this
multicritical point may be determined by noting that it corresponds to
a higher order NP root of the fixed point equations. This will be
discussed in some detail in the appendix. 

\begin{figure}[!h]
\centering
\includegraphics[width=8cm]{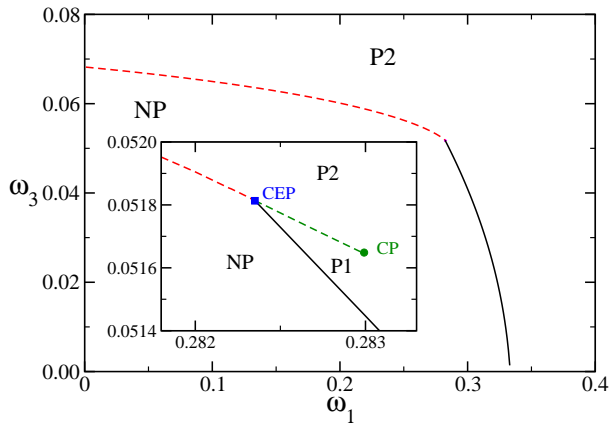}
\caption{(Color online) Phase diagram for $\omega_{2}=0$. At the red
  (left of the critical endpoint-CEP) and the magenta (right of the
  CEP) curves we find 
  first order transitions between the polymerized phase (P1) and the
  non-polymerized phase (NP) and between both polymerized phases,
  respectively. The black line (below the CEP) corresponds to continuous
  transitions between the NP and the polymerized phases. The blue (square)
  and magenta (circle) points are the critical endpoint (CEP) and a
  critical point (CP),
  respectively. All these features are better seen in the inset, which
  shows a small region close to the point where the discontinuous
  (NP-P2) and continuous (NP-polymerized) transitions lines meet.} 
\label{w1w3}
\end{figure}

A rich phase diagram was found in the $\omega_{2} = 0$ plane, as show
in Fig. \ref{w1w3}. For small values of $\omega_{1}$ there is a first
order transition between the NP em P2 phases, ending at a critical
end-point (CEP), located at $\omega_{1}^{CEP} \approx 0.2823$ and
$\omega_{3}^{CEP} \approx 0.0518$. In a tiny region of the phase
diagram, where $\omega_{1}>\omega_{1}^{CEP}$ and $\omega_{3}
\lessapprox \omega_{3}^{CEP}$, we found the second polymerized phase
(P1). The two polymerized phases (P1 and P2) coexists in line which
limits this region until 
a critical point is reached (at $\omega_{1}^{CP} \approx 0.2831$ and
$\omega_{3}^{CP} \approx 0.0516$). This is shown in the inset of figure
\ref{w1w3}. Below the coexistence line (for
$\omega_{1}>\omega_{1}^{CEP}$ and $\omega_{3} < \omega_{3}^{CEP}$),
there is a continuous transition line between the NP and the polymerized
phases. 

\begin{figure}[!h]
\centering
\includegraphics[width=8cm]{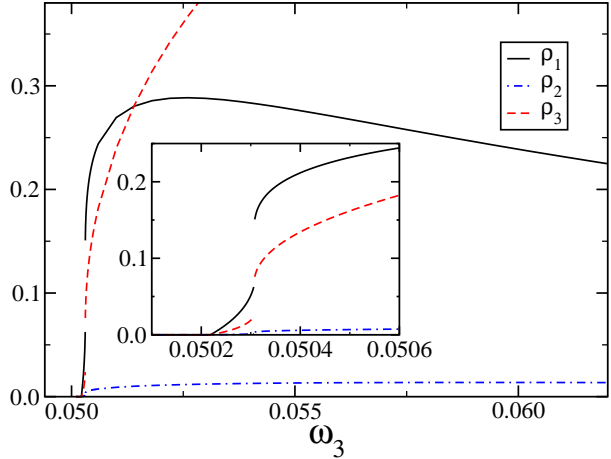}
\caption{(Color online) Densities as a function of $\omega_{3}$, for
  $\omega_{1}=0.2799$ and $\omega_{2}=0.004$. In the inset we show a
  detail of the region with discontinuous transitions.} 
\label{dens}
\end{figure}

Comparing the phase diagram of Fig. \ref{w1w3} with the one of the RF
model (for $\omega_{2}=0$) \cite{tiago09}, we can see that the
qualitative picture is the same. However, in the RF model, the
P1-P2 coexistence region is larger than the one for the RA
model. Besides, as shown in \cite{tiago09}, for the RF model, the
phase P1 is characterized by $\rho_{1}\gg \rho_{2},\rho_{3}$ at the
coexistence with the P2 phase, where $\rho_{1} \sim \rho_{2} \sim
\rho_{3}$. On the other hand, here we found that in the P1 phase all
densities are of the same order as in P2, and thus the two phases
have the same symmetries. In figure \ref{dens}, we show the densities
(defined in Eq. \ref{den}) for increasing values of $\omega_{3}$, with
$\omega_{1}$ and $\omega_{2}$ fixed in the P1-P2 coexistence
region. In face of this result, we can conclude that the restriction
imposed in the RF model changes the nature of the P1 phase, which
became approximately a SAW in that case, with dominance of sites with
a single monomer, while when immediate
reversals (RA model) for the walks are allowed,  the P1 phase behaves
like a regular polymerized phase in the MMS model. We advance that
this difference of the P1 phase in 
the two models may explain the different regimes found in the phase
boundaries in the canonical
simulations of Krawczyk et al \cite{kpor06}. This point will be
discussed in more detail below. 

The ($\omega_{1},\omega_{3}$) phase diagrams are similar to the one in
Fig. \ref{w1w3}, for all $\omega_{2}$ smaller then the multicritical
point value ($\omega_{2}<\omega_{2}^{MC}$). Therefore, we have lines of CP
and CEP in the whole phase diagram and a P1-P2 coexistence surface
between these lines. Thus, there exist a NP-P2 coexistence
surface and a NP-polymerized critical surface in the three
dimensional diagram. For increasing values of $\omega_{2}$, the CP
line gets closer to the CEP line, making the numerical
determination of their locations very hard. At the multicritical point
these lines meet, together with the TCP line. When
$\omega_{2}^{MCP} < \omega_{2} < \omega_{2}^{TCP (K=2)}$, the
tricritical point line crosses the ($\omega_{1},\omega_{3}$) plane and
the $\omega_{1} \times \omega_{3}$ diagrams resembles the one shown in
Fig. \ref{w1w2}.

\begin{figure}[!h]
\centering
\includegraphics[width=8cm]{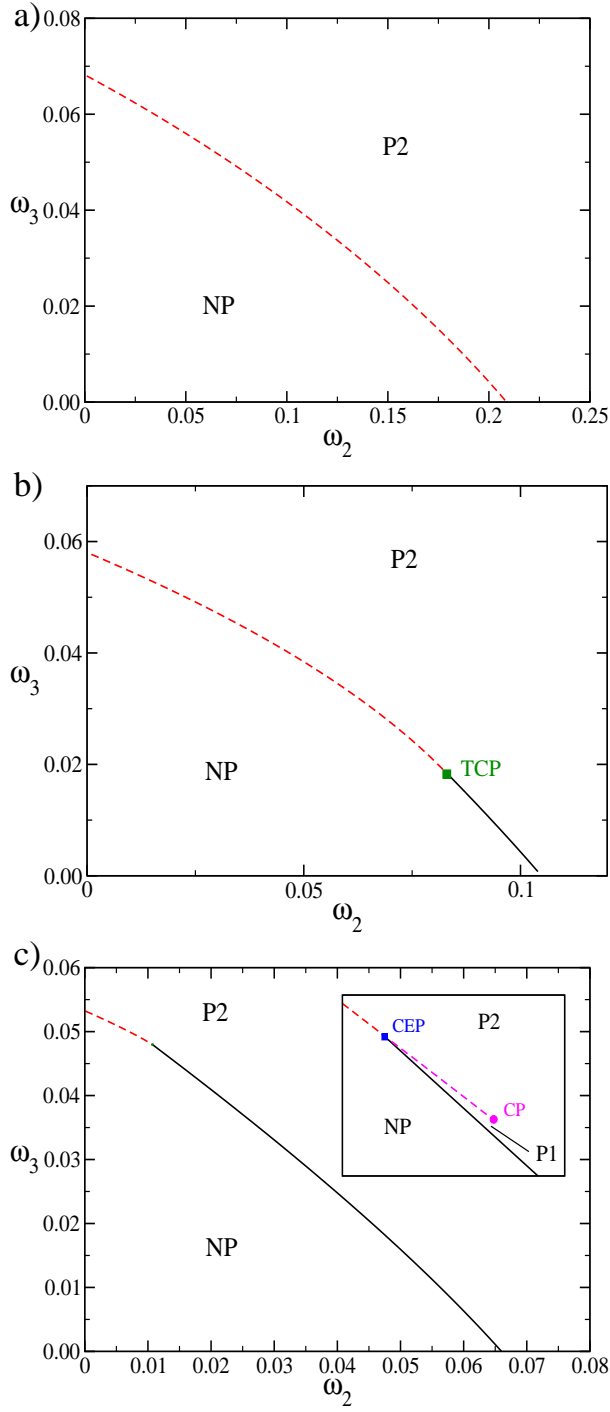}
\caption{(Color online) Phase diagrams for: a) $\omega_{1}=0.0$, b)
  $\omega_{1}=0.23$ and c) $\omega_{1}=0.275$. The red and magenta
  (dashed) curves are NP-P2 and P1-P2 discontinuous transitions,
  respectively. The black line is the NP-polymerized continuous
  transition.} 
\label{w2w3}
\end{figure}

In Fig. \ref{w2w3}, we show several diagrams, in the
($\omega_{2},\omega_{3}$) plane, for different fixed values of
$\omega_{1}$. For $\omega_{1}=0$ (Fig. \ref{w2w3}(a)), there is only a
NP-P2 coexistence line and, for $\omega_{1} < \omega_{1}^{TCP (K=2)}$,
similar diagrams are obtained, forming the NP-P2
coexistence surface. In the range $\omega_{1}^{TCP (K=2)} <
\omega_{1} < \omega_{1}^{MCP}$, the NP-P2 transition may be continuous
or discontinuous, and both transition lines meet at a tricritical
point. In Fig. \ref{w2w3} (b) we show an example of these diagrams for 
$\omega_{1}=0.23$. Finally, for $\omega_{1}>\omega_{1}^{MCP}$, we find
the same behavior of the ($\omega_{1}$, $\omega_{3}$) diagram for
small $\omega_{2}$ (see Fig. \ref{w1w3}), with two coexistence lines
(NP-P2 and P1-P2) which meet at a critical end-point, where
the line of continuous transition between the NP and the
polymerized phases ends (see Fig. \ref{w2w3} (c)). 

Again, the diagrams found here with fixed $\omega_{1}$ are similar to
those of the RF model \cite{tiago09}. The main difference is that the
tricritical and critical end-point lines of the RA model are functions
of the three parameters ($\omega_{1}$, $\omega_{2}$ and $\omega_{3}$),
while in the RF model these lines lie in the plane
$\omega_{1}=1/3$. In the same way, in the RA model the NP-polymerized
continuous transition appear as a curved surface, while in the RF
model, it is located in the plane $\omega_{1}=1/3$. 

\begin{figure}[!h]
\centering
\includegraphics[width=11cm]{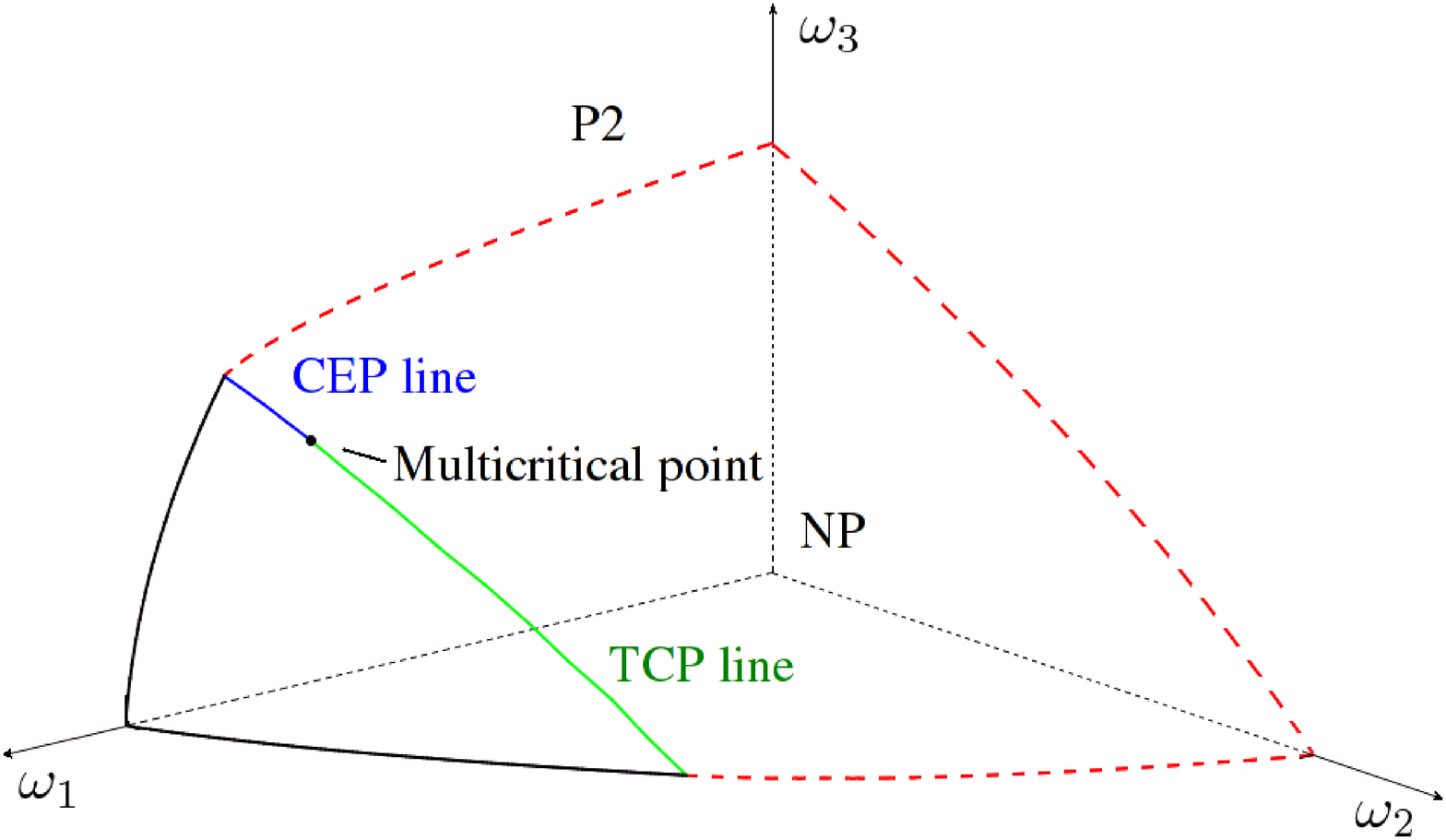}
\caption{(Color online) Sketch of the three-dimensional phase
  diagram. The NP-P2 discontinuous transition surface (red, dashed
  lines) and the NP-polymerized critical surface (black, continuous
  line) are shown. The P1-P2 coexistence surface and critical line
  close to the CEP line are to small to be represented here.} 
\label{diag3D}
\end{figure}

A sketch of the whole three-dimensional phase diagram is shown in
Fig. \ref{diag3D}, summarizing the features we have discussed
above. Like 
discussed above, the CEP and CP lines are very close in the phase
diagram and we can not see the two lines in the figure
\ref{diag3D}. Therefore, we show only the CEP line in the diagram, but
it is important to keep in mind that there is also a CP line in the 
neighborhood of this line. In particular, due to the existence of this
additional coexistence surface between both polymerized phases, the
NP-P critical surface and the NP-P coexistence surface do not meet
tangentially at the CEP line, and the angle between the normal vectors
to both surfaces at this line becomes larger as further
we are from the multicritical point, where the tricritical, critical
endpoint and P1-P2 critical line meet tangentially. Therefore, this
angle is largest when $\omega_2=0$, as may be seen in Fig. \ref{w1w3}.  

\section{Final discussions and conclusions}
\label{conclusao}

Although the Bethe lattice solution of the RA model is close to the
one of the restrictive RF model, the polymerized phases P1 are very
different in 
the two models. Moreover, in the RF model the continuous transition
surface exists between the NP and P1 phases only, while here the main
part of this surface are between the NP and a polymerized phase that
can not be identified as P1 or P2, because they are above the critical
point line. These results may explain the difference found in the
canonical simulations of Krawczyk et al \cite{kpor06} between the RA
and RF models. 
For the RF model, Krawczyk et al \cite{kpor06} suggest a canonical
phase diagram with discontinuous and continuous transition lines
which meet at a tricritical point, between a SAW-like phase (sites
occupied mainly by a single monomer) and a collapsed one (sites
predominantly with two or three monomers). The location of the TCP is
not defined precisely by the simulations, but the authors suggest that
it is in 
the region of attractive interaction between monomers, namely, the
first quadrant in the $(\beta_1,\beta_2)$ parameter space. The Bethe
lattice solution of this model 
\cite{tiago09} shows that, in fact, there is a SAW-like phase (the
critical surface NP-P1) and a collapsed phase (the coexistence surface
NP-P2). However, the discontinuous and continuous transitions lines,
suggested in the simulations, are a tricritical and a CEP line in this
approximation, respectively. 
On the other hand, no SAW-collapsed transition was found in the
simulations of the RA model in \cite{kpor06}. This is in agreement with
our results, because here the critical surface NP-polymerized does not
lead to a SAW-like phase in the canonical diagram. In contrast with
the RF model, where the sites are predominantly visited by one
monomer, here the densities in phase P1 depends only on the
statistical weights ($\omega_{i}$), like in phase P2. Thus, both the
critical surface NP-polymerized and the coexistence surface NP-P2, are
associated with collapsed phases, since the sites are occupied
predominantly by more 
than one monomer in both cases. The former leads to a collapsed phase
with low density (CLD) of monomers and the last has a larger density
(CHD). We believe that the similarity between these phases makes it
difficult to distinguish them in the simulations, which could have lead
to the conclusion of no transition for the canonical RA model in
\cite{kpor06}.  

\begin{figure}[!h]
\centering
\includegraphics[width=8cm]{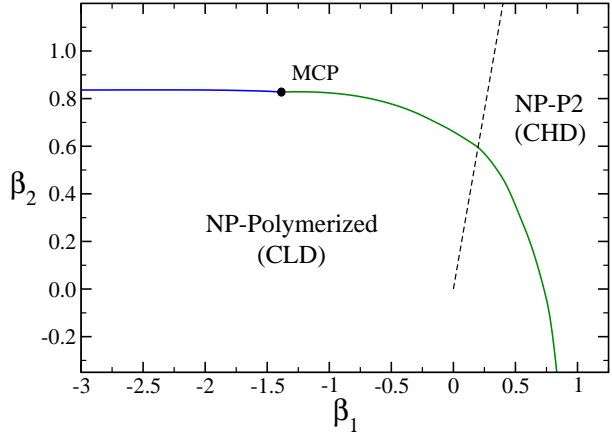}
\caption{(Color online) Canonical phase diagram. The green curve
  (below the MCP, represented by the black circle) is the tricritical
  line and the blue (above the MCP) is the CEP line. On the line
  $\beta_2=3\beta_1$ (dashed) the MMS-RA model is related to the
  SASAW's model.}  
\label{canonico}
\end{figure}

In order to compare our grand-canonical results for the RA with the
canonical ones obtained in the simulations for the RF model
\cite{kpor06}, we map our grand-canonical diagram into
canonical one. As was discussed in \cite{tiago09}, the canonical
variables used in the simulations \cite{kpor06} are related to the
Boltzmann weights of our solution as: 
\begin{equation}
\beta_{1}=\ln{\left[ \frac{\omega_{2}}{\omega_{1}^{2}}\right] } \quad
\quad \text{and} \quad \quad \beta_{2}=\ln{\left[
    \frac{\omega_{3}}{\omega_{1}^{3}}\right] },
\label{beta}
\end{equation}
and in the canonical formalism we are always restricted to the
boundary of the NP phase.
The canonical phase diagram which we found in the $\beta_{1},
\beta_{2}$ parameter space is shown in Fig. \ref{canonico}. Like
discussed 
above, we found two collapsed phases with high (CHD) and low (CLD)
monomer densities that are related to the grand-canonical NP-P2
coexistence surface and NP-polymerized critical surface, respectively.  

The CLD-CHD transitions are always continuous, but of different types:
for values of $\beta_2$ below the multicritical point there is a
tricritical line and above 
this point a critical end-point line separates the two phases. The
multicritical point is located at $\beta_{1}=-1.3840554$ and
$\beta_{2}=0.8277081$. The same behavior was found in the Bethe lattice
solution of the RF model \cite{tiago09}, but there the whole TCP line
lies in 
the region of negative values for $\beta_{2}$ negative axis
($\beta_{1}=0$, $\beta_{2}<0$) and the MCP is placed at the origin
($\beta_{1}=\beta_{2}=0$). 
Curiously, here the MCP is located in a region with repulsive interaction
between monomers at same site, when only two monomers are
present. Thus, in this region, sites occupied by two monomers are
penalized, while sites with one or three monomers are favored. In
fact, in the Fig. \ref{dens}, that show the densities in a region
close to the MCP, we can see that $\rho_{1}, \rho_{3} \gg \rho_{2}$. 

A connection can be established between the MMS and the SASAW's
models, as was already discussed in \cite{tiago09}. If we suppose that
only monomers located at the same site have through an attractive
pairwise interaction of energy $-\epsilon$, in the grand-canonical
ensemble we should have $\omega_1=z$, $\omega_2=z^2\omega$, and
$\omega_3=3z^3\omega$, where, as before,
$\omega=\exp(\beta\epsilon)$. In the canonical situation, from
Eqs. (\ref{beta}), this leads to $\beta_2=3\beta_1=3\omega$. This line
is shown in Fig. \ref{canonico}, and it crosses the tricritical line,
so that the collapse transition in the subspace of the parameter space
where the MMS-RA model on the Bethe lattice with $K=3$ is related to
the SASAW's model is tricritical, as it is also in the SASAW's model.

In conclusion, to study the thermodynamic behavior of models for
complex fluids such as the one considered here, for which exact
solutions are usually difficult to obtain, we think it is useful
to combine numerical simulations with approximate analytic
solutions. In particular, the findings in this work suggest that the
qualitative behavior of the MMS model without restrictions (RA) may 
be similar to tho one of the restricted model
(RF). Also, on the Bethe lattice, the MMS model shows a behavior which
is close to the one of the standard SAW's model for the collapse
transition of polymers. Of course one has to be aware that Bethe
lattice solutions, as all mean-field like approximations, overestimate
the interactions, and therefore may predict phase transitions in
situations where better approximations or exact solutions find none,
but in our opinion the results found on the Bethe lattice for the MMS
model suggest that additional investigations using simulations or other
more precise techniques are desirable.

\section*{Acknowledgments}

We acknowledge partial financial support by the brazilian agencies
CNPq and FAPERJ. 

\appendix

\section{Determination of the location of the multicritical point} 
\label{append}

The multicritical point may be defined as the common point of the
lines of tricritical points and critical endpoints, as may be seen in
the full phase diagram depicted in Fig. \ref{diag3D}. This definition,
however, does not lead directly to a precise algorithm to determine the
location of the multicritical point, due to the rather intricate
nonlinear fixed point equations which need to be handled for this
purpose. Sometimes, in Bethe or Husimi lattice solutions, it may be
possible to reduce the fixed point equations to a polynomial, and then
the higher order transition loci can be identified with the higher
order roots of the NP fixed point, and example of this procedure is
described  for the particular case $K=2$ of the RF-RA model in
\cite{ss10}. We were not able to pursue accomplish this calculation in
the present case. 

Another possibility, which was used for the $K=3$ RF model in
\cite{tiago09}, is to assume that, close to the NP fixed point, the
remaining ratios may be expanded in powers of one of them. By
expanding the fixed point equations in powers of the chosen ratio, one
then requires the higher order transition loci to be a higher order
root of the resulting set of equations in the parameters of the model
and the expansion coefficients. In the present case, we expanded the
remaining ratios in powers of $R_1$ and found a consistent solution of
this kind by requiring the multicritical point to satisfy the expanded
fixed point equations up to order $R_1^5$. This rather high order is
necessary due to parity effects we found in the expansions of the
ratios. We assumed that, up to order $R_1^5$, the ratios may be
expanded as follows:
\begin{subequations}
\begin{eqnarray}
R_2&=&a_{22}R_1^2+a_{24}R_1^4,\\
R_3&=&a_{30}+a_{32}R_1^2+a_{34}R_1^4,\\
R_4&=&a_{43}R_1^3+a_{45}R_1^5,\\
R_5&=&a_{51}R_1+a_{53}R_1^3+a_{55}R_1^5,\\
R_6&=&a_{62}R_1^2+a_{64}R_1^4,\\
R_7&=&a_{70}+a_{72}R_1^2+a_{74}R_1^4,\\
R_8&=&a_{81}R_1+a_{83}R_1^3+a_{85}R_1^5.
\end{eqnarray}
\label{ecr}
\end{subequations}
Now these expressions are substituted into the 8 fixed point equations
which are obtained by imposing $R_i'=R_i$ in the recursion relations
for the ratios Eqs. (\ref{rr1}-\ref{rr8}). Expanding these fixed point
equations 
up to order $R_1^5$, using a algebra software for this task, we obtain
several expansion coefficients of the fixed point
equations. Considering the parity of the expansions shown in
Eqs. (\ref{ecr}) and denoting by $C_{ij}$ the coefficient of
order $R_1^j$ in the fixed point equation obtained from the recursion
relation for $R_i$, we are lead to 21 equations $C_{i,j}=0$,
corresponding the the coefficients $(1,1)$, $(1,3)$, $(1,5)$, $(2,2)$, 
$(2,4)$, $(3,0)$, $(3,2)$, $(3,4)$, $(4,3)$, $(4,5)$, $(5,1)$,
$(5,3)$, 
$(5,5)$, $(6,1)$, $(6,3)$, $(6,5)$, $(7,0)$, $(7,2)$, $(7,4)$, $(8,1)$,
$(8,3)$ 
and $(8,5)$. In particular, $C_{3,0}=0$ and $C_{7,0}=0$ lead to the
pair of equations (\ref{npeq1},\ref{npeq2}) for the fixed point values
of the NP phase presented above. The complete set of equations is too
long to be given here. Solving this set of nonlinear algebraic
equations  for the 18 expansion coefficients $a_{ij}$ and the 3
statistical weights $\omega_i$, we obtain the result: $a_{22} =
0.4882335108$,  $a_{24} = 0.2869523407$, $a_{30} = 0.3329767741$,
$a_{32} = -0.03809592295$, $a_{34} = 0.1519415331$, $a_{43} =
0.2048247034$, $a_{45} = -0.2633449211$, $a_{51} = 0.3220483966$,
$a_{53} = 0.1813751089$, $a_{55} = -0.6519478195$, $a_{62} =
0.2012557404$, $a_{64} = -0.2612169532$, $a_{70} = 0.1083347046$,
$a_{72} = 0.03872548434$, $a_{74} = -0.1169008783$, $a_{81} =
0.1352312919$, $a_{83} = -0.2015959772$, and $a_{85} = 0.1269313668$
for the expansion coefficients and $\omega_1 = 0.2700819945$,
$\omega_2 = 0.01827694593$, and $\omega_3 = 0.04507709731$ for the
statistical weights. We did some numerical calculations for the fixed
point values of the ratios in the neighborhood of the multicritical
point, and found that they are consistent with the expansion
coefficients we found.

\end{document}